\documentclass[twocolumn]{aastex631}
\usepackage{csvsimple}
\usepackage{multirow}
\usepackage{rotating}
\usepackage{amsmath}
\usepackage{longtable}
\usepackage{booktabs}
\usepackage{ltablex}
\usepackage{epsfig}
\usepackage{threeparttable}
\usepackage{url}

\def\xmm{\emph{XMM--Newton}}
\def\cxo{\emph{Chandra}}

\def\swift{\emph{Swift}}
\def\nicer{\emph{NICER}}
\def\ep{\emph{Einstein Probe}}

\def\flux{\mbox{erg\,cm$^{-2}$\,s$^{-1}$}}
\def\lum{\mbox{erg\,s$^{-1}$}}

\def\cm2{\rm \ cm$^{-2}$}

\def\arcmin{\mbox{$^{\prime}$}}

\def\srcfirst{EP\,J005245.1$-$722843}
\def\srcalt{CXOU\,J005245.0$-$722844}

\def\src{J0052}

\newcommand\T{\rule{0pt}{2.6ex}}       
\newcommand\B{\rule[-1.2ex]{0pt}{0pt}}

\shorttitle{Discovery of a rare BeWD binary in the SMC}

\begin{document}

\title{Einstein Probe discovery of EP\,J005245.1$-$722843:\\ a rare BeWD binary in the Small Magellanic Cloud?}

\correspondingauthor{A.~Marino, H.~N.~Yang}
\author[0000-0001-5674-4664]{A.~Marino}
\email{marino@ice.csic.es, hnyang@nao.cas.cn}
\affiliation{Institute of Space Sciences (ICE, CSIC), Campus UAB, Carrer de Can Magrans s/n, 08193, Barcelona, Spain}
\affiliation{Institut d'Estudis Espacials de Catalunya (IEEC), Esteve Terradas 1, RDIT Building, Of. 212 Mediterranean Technology Park (PMT), 08860, Castelldefels, Spain} 
\author[0000-0002-7680-2056]{H.~N.~Yang}
\affiliation{National Astronomical Observatories, Chinese Academy of Sciences, Beijing, 100101, People's Republic of China}
\affiliation{Max-Planck-Institut f\"{u}r extraterrestrische Physik, Gie\ss{}enbachstra\ss{}e 1, D-85748 Garching bei M\"{u}nchen, Germany}
\affiliation{School of Astronomy and Space Sciences, University of Chinese Academy of Sciences, Beijing, 100049, People's Republic of China}
\author[0000-0001-7611-1581]{F. Coti Zelati}
\affiliation{Institute of Space Sciences (ICE, CSIC), Campus UAB, Carrer de Can Magrans s/n, 08193, Barcelona, Spain}
\affiliation{Institut d'Estudis Espacials de Catalunya (IEEC), Esteve Terradas 1, RDIT Building, Of. 212 Mediterranean Technology Park (PMT), 08860, Castelldefels, Spain} 
\author[0000-0003-2177-6388]{N.~Rea}
\affiliation{Institute of Space Sciences (ICE, CSIC), Campus UAB, Carrer de Can Magrans s/n, 08193, Barcelona, Spain}
\affiliation{Institut d'Estudis Espacials de Catalunya (IEEC), Esteve Terradas 1, RDIT Building, Of. 212 Mediterranean Technology Park (PMT), 08860, Castelldefels, Spain} 
\author[0000-0002-6449-106X]{S.~Guillot}
\affiliation{IRAP, CNRS, 9 avenue du Colonel Roche, BP 44346, 31028 Toulouse Cedex 4, France}
\author[0000-0002-6789-2723]{G.~K. Jaisawal}
\affiliation{DTU Space, Technical University of Denmark, Elektrovej 327-328, DK-2800 Lyngby, Denmark}
\author[0000-0002-0766-7313]{C.~Maitra}
\affiliation{Max-Planck-Institut f\"{u}r extraterrestrische Physik, Gie\ss{}enbachstra\ss{}e 1, D-85748 Garching bei M\"{u}nchen, Germany}
\author[0000-0003-0440-7193]{J.-U.~Ness} 
\affiliation{European Space Agency (ESA), European Space Astronomy Centre (ESAC), Camino Bajo del Castillo s/n, E-28692 Villanueva de la Cañada, Madrid, Spain}
\author[0000-0002-0107-5237]{F.~Haberl}
\affiliation{Max-Planck-Institut f\"{u}r extraterrestrische Physik, Gie\ss{}enbachstra\ss{}e 1, D-85748 Garching bei M\"{u}nchen, Germany}
\author[0000-0002-5790-7290]{E.~Kuulkers}
\affiliation{European Space Agency (ESA), European Space Research and Technology Centre (ESTEC), Keplerlaan 1, 2201 AZ Noordwijk,
The Netherlands}
\author{W.~Yuan}
\affiliation{National Astronomical Observatories, Chinese Academy of Sciences, Beijing, 100101, People's Republic of China}
\affiliation{School of Astronomy and Space Sciences, University of Chinese Academy of Sciences, Beijing, 100049, People's Republic of China}
\author{H.~Feng}
\affiliation{Key Laboratory of Particle Astrophysics, Institute of High Energy Physics, Chinese Academy of Sciences, \\Beijing 100049, People's Republic of China}
\author{L.~Tao}
\affiliation{Key Laboratory of Particle Astrophysics, Institute of High Energy Physics, Chinese Academy of Sciences, \\Beijing 100049, People's Republic of China}
\author{C.~Jin}
\affiliation{National Astronomical Observatories, Chinese Academy of Sciences, Beijing, 100101, People's Republic of China}
\affiliation{School of Astronomy and Space Sciences, University of Chinese Academy of Sciences, Beijing, 100049, People's Republic of China}
\author[0000-0002-9615-1481]{H.~Sun}
\affiliation{National Astronomical Observatories, Chinese Academy of Sciences, Beijing, 100101, People's Republic of China}
\author{W.~Zhang}
\affiliation{National Astronomical Observatories, Chinese Academy of Sciences, Beijing, 100101, People's Republic of China}
\author{W.~Chen}
\affiliation{National Astronomical Observatories, Chinese Academy of Sciences, Beijing, 100101, People's Republic of China}
\affiliation{School of Astronomy and Space Sciences, University of Chinese Academy of Sciences, Beijing, 100049, People's Republic of China}
\author{E.~P.~J.~van den Heuvel} 
\affiliation{Anton Pannekoek Institute for Astronomy, University of Amsterdam, PO Box 94249, 1090 GE Amsterdam, The Netherlands}
\author[0000-0002-4622-796X]{R.~Soria}
\affiliation{INAF – Osservatorio Astrofisico di Torino, Strada Osservatorio 20, I-10025 Pino Torinese, Italy}
\affiliation{Sydney Institute for Astronomy, School of Physics A28, The University of Sydney, Sydney, NSW 2006, Australia}
\affiliation{College of Astronomy and Space Sciences, University of the Chinese Academy of Sciences, Beijing 100049, People's Republic of China}
\author[0000-0002-9725-2524]{B.~Zhang}
\affiliation{Nevada Center for Astrophysics, University of Nevada Las Vegas, NV 89154, USA}
\affiliation{Department of Physics and Astronomy, University of Nevada Las Vegas, NV 89154, USA}
\author{S-S.~Weng}
\affiliation{Department of Physics and Institute of Theoretical Physics, Nanjing Normal University, Nanjing, People's Republic of China}
\author{L.~Ji}
\affiliation{School of Physics and Astronomy, Sun Yat-sen University, Zhuhai, 519082, People's Republic of China}
\author{G.~B.~Zhang}
\affiliation{Yunnan Observatories, Chinese Academy of Sciences, Kunming 650216, People's Republic of China}
\author{X.~Pan}
\affiliation{National Astronomical Observatories, Chinese Academy of Sciences, Beijing, 100101, People's Republic of China}
\author{Z.~Lv}
\affiliation{National Astronomical Observatories, Chinese Academy of Sciences, Beijing, 100101, People's Republic of China}
\affiliation{School of Astronomy and Space Sciences, University of Chinese Academy of Sciences, Beijing, 100049, People's Republic of China}
\author{C.~Zhang} 
\affiliation{National Astronomical Observatories, Chinese Academy of Sciences, Beijing, 100101, People's Republic of China}
\affiliation{School of Astronomy and Space Sciences, University of Chinese Academy of Sciences, Beijing, 100049, People's Republic of China}
\author{Z.~X.~Ling}
\affiliation{National Astronomical Observatories, Chinese Academy of Sciences, Beijing, 100101, People's Republic of China}
\affiliation{School of Astronomy and Space Sciences, University of Chinese Academy of Sciences, Beijing, 100049, People's Republic of China}
\author{Y.~Chen} 
\affiliation{Key Laboratory of Particle Astrophysics, Institute of High Energy Physics, Chinese Academy of Sciences, \\Beijing 100049, People's Republic of China}
\author{S.~Jia} 
\affiliation{Key Laboratory of Particle Astrophysics, Institute of High Energy Physics, Chinese Academy of Sciences, \\Beijing 100049, People's Republic of China}
\affiliation{School of Astronomy and Space Sciences, University of Chinese Academy of Sciences, Beijing, 100049, People's Republic of China}
\author{Y.~Liu}
\affiliation{National Astronomical Observatories, Chinese Academy of Sciences, Beijing, 100101, People's Republic of China}
\author{H.~Q.~Cheng}
\affiliation{National Astronomical Observatories, Chinese Academy of Sciences, Beijing, 100101, People's Republic of China}
\author{D.~Y.~Li}
\affiliation{National Astronomical Observatories, Chinese Academy of Sciences, Beijing, 100101, People's Republic of China}
\author[0000-0001-7115-2819]{K.~Gendreau}
\affiliation{Astrophysics Science Division, NASA's Goddard Space Flight Center, Greenbelt, MD 20771, USA}
\author[0000-0002-0940-6563]{M.~Ng} 
\affiliation{MIT Kavli Institute for Astrophysics and Space Research, Massachusetts Institute of Technology, Cambridge, MA 02139, USA}
\author[0000-0001-7681-5845]{T.~Strohmayer} 
\affiliation{Astrophysics Science Division, NASA's Goddard Space Flight Center, Greenbelt, MD 20771, USA}
\affiliation{Joint Space-Science Institute, NASA Goddard Space Flight Center, Greenbelt, MD 20771, USA}


\received{July, 2024}
\submitjournal{ApJL}

\begin{abstract}
On May 27 2024, the Wide-field X-ray Telescope onboard the \ep\ (EP) mission detected enhanced X-ray emission from a new transient source in the Small Magellanic Cloud (SMC) during its commissioning phase. Prompt follow-up with the EP Follow-up X-ray Telescope, the \swift\ X-ray Telescope and \nicer\ have revealed a very soft, thermally emitting source (kT$\sim$0.1\,keV at the outburst peak) with an X-ray luminosity of $L\sim4\times10^{38}$\,erg~s$^{-1}$, labelled \srcfirst. This super-soft outburst faded very quickly in a week time. Several emission lines and absorption edges were present in the X-ray spectrum, including deep Nitrogen (0.67\,keV) and Oxygen (0.87\,keV) absorption edges. The X-ray emission resembles the SSS phase of typical nova outbursts from an accreting white dwarf (WD) in a binary system, despite the X-ray source being historically associated with an O9-B0e massive star exhibiting a 17.55\,days periodicity in the optical band. The discovery of this super-soft outburst suggests that \srcfirst\ is a BeWD X-ray binary: an elusive evolutionary stage where two main-sequence massive stars have undergone a common envelope phase and experienced at least two episodes of mass transfer. In addition, the very short duration of the outburst and the presence of Ne features hint at a rather massive, i.e., close to the Chandrasekhar limit, Ne-O WD in the system.
\end{abstract}



\section{Introduction} \label{sec:intro}

Be-white dwarf X-ray binary systems (BeWDs) are a subclass of X-ray binaries where a white dwarf (WD) accretes matter from a Be-type main-sequence star. Binary evolution models predict that BeWDs should be about seven times more common than Be-neutron star (BeNS) systems \citep{raguzova01}. Population synthesis models \citep{zhu23} indicate that BeWDs originate from binaries initially composed of a Be star and a subdwarf O or B star. A substantial fraction of these systems can evolve either into red giants (via merging of the WD with a non-degenerated star; 60--70\% of the cases), or into double WDs detectable by the LISA mission as sources of gravitational wave emission (30--40\%). However, BeWDs have been so far relatively elusive. On the one hand, $\gamma$ Cas analogs, a class of Be stars showing an excess in hard X-rays, are largely considered to be BeWD binaries, although the matter is still controversial \citep[see, e.g.][and references therein]{Gies2023}. On the other hand, a few confirmed Be binaries which show instead rather soft X-ray outbursts, have also been identified as BeWD  \citep{Kahabka2006, Sturm2012,coe20,Haberl2020_Atel,kennea21}. The emission from the latter sources in outburst is typically characterized by very soft X-ray spectra with no detectable X-ray emission above $>$2\,keV, indicative of nuclear burning on the WD surface. In this regime, BeWDs can be classified as Super-Soft Sources (SSS) \citep{Cracco2018}, a group of sources that includes various types of X-ray systems powered by thermonuclear burning onto the surface of an accreting WD \citep{Greiner1991}. This category also encompasses novae during their SSS phase \citep[e.g.][]{Chomiuk2021}. The lack of sensitive wide-field soft X-ray monitors has made the detection of outbursts from BeWDs rather challenging, especially during the rising phase.
On May 27th 2024, the Wide-field X-ray Telescope (WXT) onboard the \ep\ (EP) mission detected enhanced X-ray emission from a new transient source, \srcfirst\, 
 (\src\ hereafter), in the SMC during its commissioning phase \citep{YangHN24}. This source was promptly identified as \srcalt , following a serendipitous 52-s observation of the field by the \emph{Neil Gehrels Swift Observatory} (\swift) X-ray Telescope (XRT) a few hours later \citep{Kennea2024, Gaudin2024}, as part of the \swift\ Small Magellanic Cloud Survey (S-CUBED; \citealt{kennea18}). Two follow-up NICER observations on May 28th and 29th revealed a rapid decrease in the X-ray count rate between the two observations \citep{Jaisawal2024}. The source has a bright optical counterpart (average magnitude of G$\sim$14.9), which has been first classified as an O9-B0e star by \citep{antoniou2009} and later suggested to belong to the B0 class based also on its spectrum \citep{Sheets2013,Maravelias2014,Gaudin2024}. The companion star has an estimated age of $\sim$40 Myears, according to the age of the OB population in that region of the SMC \citep{Antoniou2019}.  The long-term optical light curve shows a periodicity at $\sim$17.55 days, which is likely related to the orbital period of the system \citep{sarraj2012,treiber2024}. This Letter reports on the classification of this source as a BeWD following the discovery of its super-soft outburst by EP and its X-ray evolution. In the following, we adopt the position listed in the \cxo\ Source Catalog Release 2.1 \citep{evans2010}, R.A. = 00$^\mathrm{h}$52$^\mathrm{m}$45$\fs$13, decl. = --72$^{\circ}$28$^{\prime}$43$\farcs$87 (J2000.0) and a distance of 62\,kpc \citep{graczyk2020}. All parameter uncertainties are quoted at a 1$\sigma$ confidence level unless otherwise stated.

\begin{table*}
\caption{
\label{tab:log}
Journal of the 2024 X-ray observations of \src\ presented in this work.}
\centering
\begin{tabular}{lcccc}
\hline\hline
Telescope/Instrument (setup)	& Obs ID 		&Start -- End time (UTC)	   & Exposure & Epochs$^\dagger$ \\
					        &			& Mmm DD hh:mm:ss		& (ks)    \\
\hline
EP/WXT                      & 13600006246     & May 25 11:06:46 -- May 27 06:28:12   & 67.9 \\
EP/WXT                      & 13600006247     & May 27 08:02:48 -- May 27 09:39:06   & 2.0 \\
EP/WXT                      & 13600006248     & May 27 09:39:06 -- May 29 03:22:17   & 63.0 \\
\swift/XRT (PC)	            & 03111173076	  & May 27 22:29:57 -- May 27 22:30:50   & 0.052 \\
EP/WXT                      & 08500000102     & May 28 04:51:10 -- May 28 06:26:53   & 1.6 \\
EP/FXT (PW)                 & 08500000102     & May 28 05:31:05 -- May 28 06:23:21   & 1.5 & 1 \\
\swift/XRT (WT)	            & 00033745002	  & May 28 15:40:13 -- May 28 17:19:37   & 1.9 & 2 \\
\nicer/XTI                   & 7204550101      & May 28 16:56:06 -- May 28 23:24:20   & 4.1 & 3,4,5,6 \\
\nicer/XTI                   & 7204550102      & May 29 11:26:10 -- May 30 00:00:00   & 6.0 & 7,8,9,10,11,12 \\
\nicer/XTI                   & 7204550103      & May 30 01:21:10 -- May 30 23:22:00   & 8.0 & 13,14,15,16 \\
\nicer/XTI                   & 7204550104      & May 31 00:33:30 -- May 31 21:01:21   & 6.4 & 17,18 \\
\nicer/XTI                   & 7204550105      & Jun 1 04:32:49  -- Jun 1 17:16:20    & 1.9 & 19 \\
\nicer/XTI                   & 7204550106      & Jun 2 05:09:40  -- Jun 2 17:59:00    & 5.8 & 20 \\
\nicer/XTI                   & 7204550107      & Jun 2 23:46:40  -- Jun 3 23:09:40    & 3.4 & 21 \\
\nicer/XTI                   & 7204550108      & Jun 4 00:50:20  -- Jun 4 23:59:20    & 1.9 \\
EP/FXT (FF)                 & 08500000107     & Jun 4 12:15:38  -- Jun 4 16:19:28    & 9.1 \\
\nicer/XTI                   & 7204550109      & Jun 5 10:33:20  -- Jun 5 13:49:20    & 1.3 \\
\nicer/XTI                   & 7204550110      & Jun 6 02:24:40  -- Jun 6 22:23:00    & 0.8 \\
\nicer/XTI                   & 7204550111      & Jun 6 23:43:40  -- Jun 7 21:49:40    & 8.5 \\
\nicer/XTI                   & 7204550112      & Jun 8 10:14:20  -- Jun 8 18:03:00    & 0.3 \\
\swift/XRT (PC)	            & 00033745004     & Jun 13 20:35:38 --  Jun 13 22:32:52  & 2.5 \\
\xmm/EPIC-pn (SW)	        & 0935191301 	  & Jun 14 19:12:05 -- Jun 15 03:59:07   & 31.6	\\
\xmm/EPIC-MOS1 (T)	        & 0935191301	  & Jun 14 19:26:34 -- Jun 15 03:54:49   & 27.4	\\
\xmm/EPIC-MOS2 (T)	        & 0935191301 	  & Jun 14 19:26:51 -- Jun 15 03:54:39   & 29.4	\\
\swift/XRT (PC)	            & 00033745005     & Jun 15 02:31:09 --  Jun 15 21:52:53  & 1.6 \\
\swift/XRT (PC)	            & 00033745006     & Jun 17 19:09:39 --  Jun 17 21:12:52  & 2.6 \\
\swift/XRT (PC)	            & 00033745007     & Jun 19 18:35:06 --  Jun 19 18:56:53  & 1.3 \\
\hline
\hline
\end{tabular}
\newline
{\bf Notes.} FF: full frame; PW: Partial Window; PC: Photon Counting; WT: Windowed Timing; SW: Small Window; T: Timing; $^\dagger$: the observations considered for the spectral analysis in this paper. 
\end{table*}

\section{Observations and data reduction}
In this paper, we have used observations from various X-ray telescopes, including the WXT and Follow-up X-ray Telescope (FXT) onboard EP, \swift, \nicer\ and \xmm, as well as archival \cxo\ observations (see Table\,\ref{tab:log} for details). 

\subsection{\ep\,}\label{sec:ep}
As a space observatory to discover and characterize X-ray transients, EP \citep{Yuan22} carries two instruments, the WXT and FXT. EP was launched on January 9, 2024 into a low-Earth circular orbit of 592 km height with an orbital period of 97 minutes.
The WXT employs Lobster-eye micro-pore optics (MPO) to provide an instantaneous field-of-view of $\sim$3600 square degrees. The instrument comprises 12 modules, each covering about 300 square degrees. Each module includes a mirror assembly with 36 MPO optics, an optical baffle, and a focal-plane detector array of CMOS sensors.
The in-orbit calibrations of WXT have been completed, whose results are in excellent agreement with those of the on-ground calibrations (Cheng H., in preparation).
\src\ was simultaneously detected by the CMOS detectors 14 and 37 of the WXT during observations on 2024 May 27 from 08:41:28 to 12:47:16 UTC. Data were processed using the \texttt{wxtpipeline} tool version 0.1.0. The source region was determined as a circle with a radius of 67 pixels (1 pixel $\sim$ 0.136\arcmin) centered on the source. In the field of view of WXT, there are multiple bright objects together with their cross-arms. Therefore, we carefully selected a circle of radius 268 pixels nearby as the background region to minimize the influence of other bright objects.

The FXT was configured in partial window mode and full frame mode for the observations performed on May 28 and June 4, respectively. Data were processed using the \texttt{fxtchain} tool available within the FXT Data Analysis Software (\texttt{FXTDAS}). Source and background photons were extracted using a circle of radius 70\arcsec\ centered on the source and a closeby circle of radius 128\arcsec, respectively. 

\subsection{\swift}\label{sec:swift}
The XRT observations were processed using \texttt{xrtpipeline}, employing the latest version of CALDB (v.\,\texttt{20240522}). We extracted photons using a 20\arcsec\ circular region centered on the source. In the first observation (WT mode), we estimated the background with a similar-sized region far from the source. In the second observation (PC mode), we used an annular region with inner and outer radii of 45\arcsec\ and 150\arcsec\ for background estimation. We then extracted light curves and spectra through the \texttt{xrtproducts} pipeline. For the last four observations, we combined the data (totaling 8\,ks of exposure) and set a 3$\sigma$ upper limit on the net count rate of 0.002\,counts\,s$^{-1}$ (calculated over the 0.3--10\,keV energy range using the \texttt{sosta} command). 
In the first two \swift\ observations considered in this work, the UVOT was operated in image mode, using the UVW1 filter in the first one and cycling through all available filters in the second one. In the last four observations, the UVOT was operated in event mode using the UVM2 filter. We used a circle with a radius of 5\arcsec\ radius for the source and a nearby circle with a radius of 10\arcsec\ devoid of optical sources for the background. For image mode data, we performed aperture photometry (properly corrected to infinite aperture) with \texttt{uvotsource}.  
The optical brightness in the UVM2 filter varied from $\sim$12.6 to 13.2 mag (Vega) between May 27 and June 19, without showing any clear monotonic trend. For event mode data, we processed the files with \texttt{coordinator} and \texttt{uvotscreen} and searched for periodic signals using the unbinned event files (see Sect.\,\ref{sec:periodicity}).

\subsection{\nicer}\label{sec:nicer}
\nicer\, data were processed using the \texttt{nicerl2} pipeline with default screening settings \citep[following the same prescriptions as in, e.g.][]{Marino2023}. Several \nicer\ light curves showed irregular count rate increases during the outburst. To minimize charged particle contamination, we reprocessed the data with \texttt{nicerl2}, setting \texttt{overonly\_range} to 0–5 and \texttt{COR\_RANGE} to $>1.5$ GeV/$c$. Any observation segment still affected by non X-ray flares after this procedure was neglected from the following analysis. The \texttt{nicerl3-spect} pipeline was used to produce spectra and background files. In particular, the \texttt{scorpeon} model\footnote{\url{https://heasarc.gsfc.nasa.gov/docs/nicer/analysis_threads/scorpeon-overview/}} with default settings was adopted for the creation of the background file.

\subsection{\xmm}\label{sec:xmm}
The EPIC-pn onboard \xmm\ was configured in the small window mode, while both EPIC-MOS cameras operated in timing mode.
Data were processed using the Science Analysis System (SAS). We applied intensity filters to the time series of the entire field of view in the EPIC-pn data to remove several background flaring episodes (which were also visible at low energies), resulting in a net on-source exposure time of $\sim$15\,ks. No X-ray emission was detected at the source position. Using the \texttt{eupper} tool and assuming a circular region with a radius of 10 arcsec for the source photons along with an annulus with inner and outer radii of 10 and 20\arcsec\ respectively for the background, we estimated a 3-$\sigma$ upper limit on the net count rate of 0.007\,counts\,s$^{-1}$ over the 0.3--10\,keV energy range. Given the non-detection in the EPIC-pn data, we did not perform further analysis on the EPIC-MOS and RGS data.

\subsection{\cxo}\label{sec:cxo}
The field of \src\ was observed with the ACIS-I instrument on board the \cxo\ X-ray Observatory three times on July 20, 2002 (7.7\,ks), April 25, 2006 (49.3\,ks), and April 26, 2006 (47.4\,ks). 
We processed the data using the \texttt{CIAO} software package. 
In 2002, \src\ was detected at a count rate of $(1.9_{-0.8}^{+1.0}) \times 10^{-3}$\,counts\,s$^{-1}$ (0.3–7\,keV) at the 90\% c.l., corresponding to a detection significance exceeding 99\%. For the stacked observations from 2006, the count rate was $(1.3_{-0.7}^{+0.6}) \times 10^{-4}$\,counts\,s$^{-1}$ (0.3--7\,keV), representing a $\sim$3$\sigma$ detection.

\section{Data analysis and results}
\subsection{The EP/WXT light curve}
\begin{figure}
\centering
\includegraphics[scale=0.45]{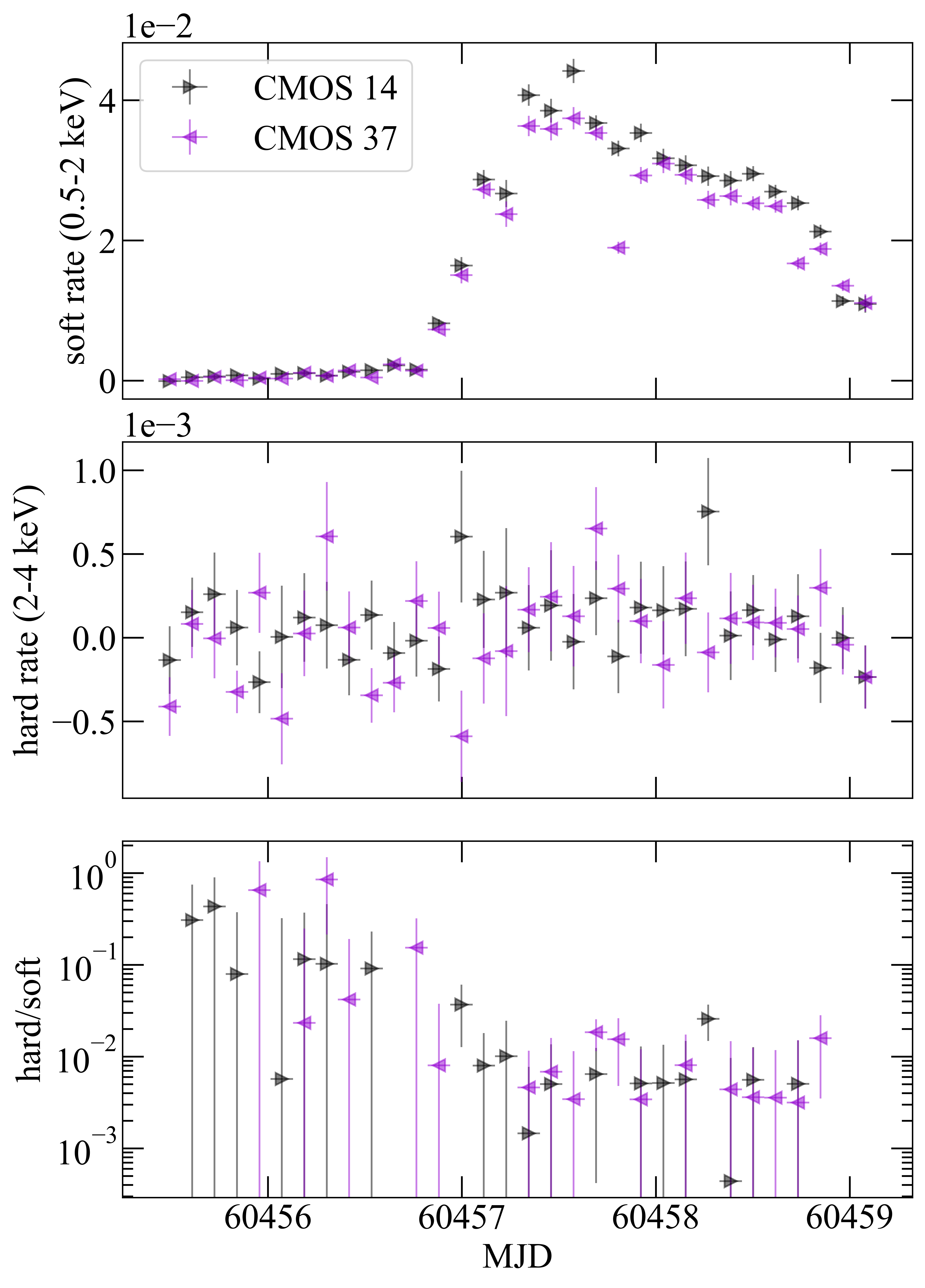}
\caption{EP/WXT light curves of \src\ around the epoch of the outburst detection in the soft (0.5--2\,keV; top) and hard (2--4\,keV; middle) energy bands. The drop in the middle of the light curve of CMOS 37 is caused by instrumental effects. The time evolution of the hardness ratio is also plotted (bottom).} 
\label{Fig:wxt_lc}
\end{figure}

The WXT light curve is shown in Figure \ref{Fig:wxt_lc}. After removing spurious drops in count rates caused by instrumental artifacts, the analysis of the observations before and after the source detection revealed a rapid increase in emission starting around MJD 60456.8, with a linear rise at $\sim$0.3\,counts\,s$^{-1}$\,day$^{-1}$, peaking around MJD 60457.5 after approximately 0.6 days. This was followed by a slow linear decay at $\sim$0.02\,counts\,s$^{-1}$\,day$^{-1}$ until around MJD 60458.8, and then a faster decay at $\sim$0.23\,counts\,s$^{-1}$\,day$^{-1}$.

\subsection{X-ray spectral analysis}
\label{sec:spectral}
\begin{figure*}
\centering
\includegraphics[scale=0.54]{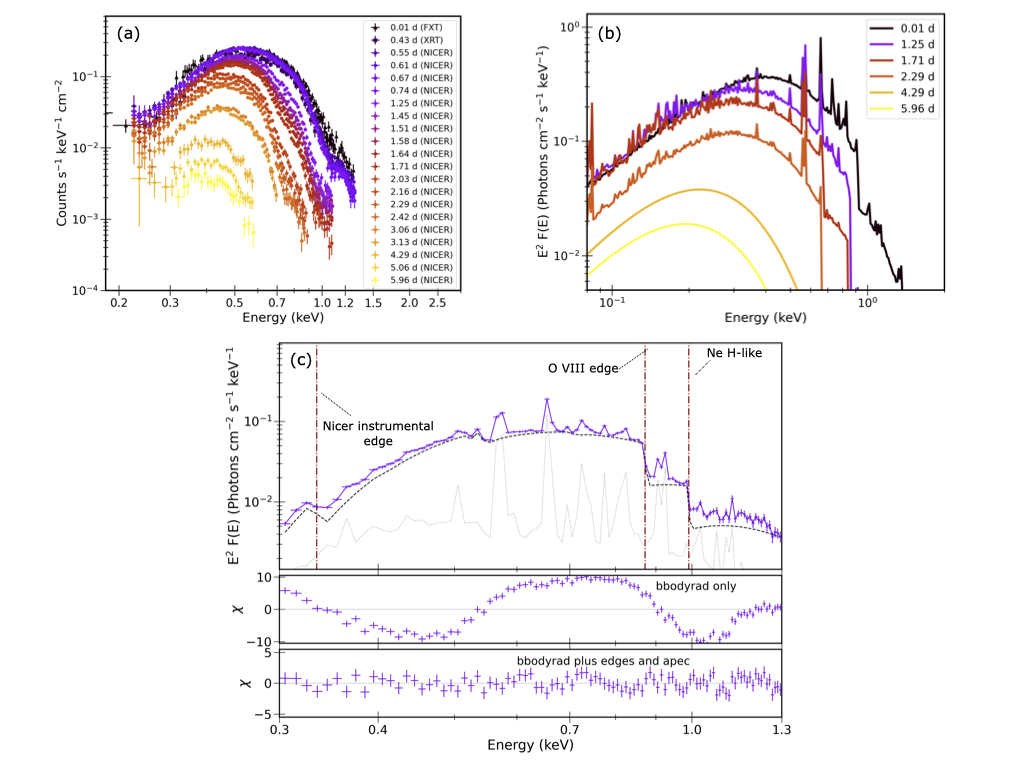}
\caption{Summary plot for the spectral analysis performed for \src. Panel (a): single spectra divided by each instrument's effective area taken by EP/FXT (triangles), \swift/XRT (squares) and \nicer\ (circles). The time for each spectrum is referenced from MJD 60458.23. Panel (b): best-fit unabsorbed models of selected observations. Panel (c): unfolded \nicer\ spectrum taken during epoch 4 along with the best-fit model (top panel). The \texttt{bbodyrad} component is represented by a dashed black line, while the \texttt{apec} component is drawn with dotted silver line. The energies of the identified edges are marked with vertical maroon lines. The
unmodelled and modelled residuals are shown in the middle and bottom panel.} 
\label{Fig:spectra}
\end{figure*}

The X-ray flux and spectral shape of \src\ varied significantly over hours. Therefore, we extracted time-resolved spectra over intervals with approximately constant count rates. For \nicer\ observations 7204550101, 7204550102, 7204550103, and 7204550104, we divided them into 4, 6, 4, and 2 epochs, respectively, extracting one spectrum for each. The signal was largely or completely background-dominated in all \nicer, EP/FXT and \swift/XRT observations performed after June 2, so we excluded these spectra from further analysis. The final spectral sample consists therefore of 1 FXT, 1 XRT and 19 \nicer\ time-resolved spectra, hereafter referred to as epochs (see Table \ref{tab:log} for a summary). We grouped all the spectra ensuring a minimum of 25 counts per bin, allowing the use of the $\chi^2$-statistics. Figure\,\ref{Fig:spectra} (a) shows the evolution of the extracted spectra of the source. We included data only where the source emission was above the background, which varied from spectrum to spectrum but was always within the 0.3--1.3 keV range. We note that despite the not-optimal calibration of \nicer\ data below $\sim$0.4\,keV\footnote{\url{https://heasarc.gsfc.nasa.gov/docs/nicer/analysis_threads/cal-recommend/}}, we decided to retain data between 0.3 and 0.4\,keV due to the source bright soft emission.

\begin{figure*}
\centering
\includegraphics[scale=0.36]{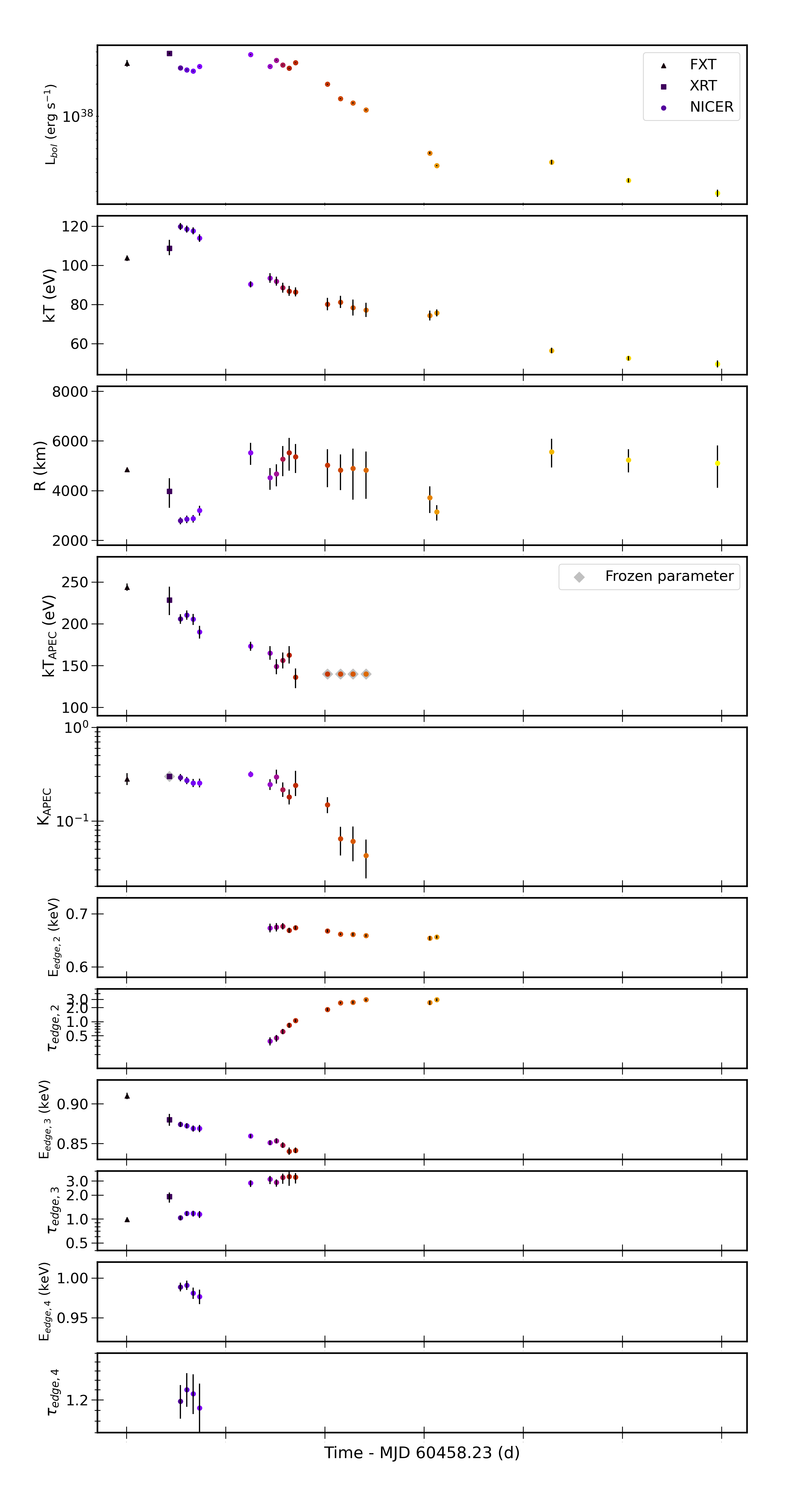}
\vspace{-0.5cm}
\caption{Best-fit spectral parameters obtained in this work and their evolution during the outburst. The color sequence is the same as the one used in Figure \ref{Fig:spectra}. See also Tables \ref{tab:continuum} and \ref{tab:discrete}.} 
\label{fig:best-fit}
\end{figure*}

We performed the spectral analysis with \texttt{xspec} v.12.14.b \citep{arnaud96}. A systematic error of 1\% was included, as recommended by the \nicer\ team\footnote{\url{https://heasarc.gsfc.nasa.gov/docs/nicer/analysis_threads/spectrum-systematic-error/}}. The interstellar absorption was modelled through the \texttt{TBabs} component, with abundances from \citet{Wilms2000} and cross sections from \citet{Verner1996}. We derived the unabsorbed bolometric fluxes between 0.001 and 10\,keV using the \texttt{cflux} convolution model. We initially adopted a simple absorbed blackbody model (\texttt{BBodyrad}) to fit the spectra. Although it matched the continuum reasonably well, the overall fit was poor, with significant unmodeled residuals (see, e.g., Figure\,\ref{Fig:spectra}, panel c). For \nicer, artifacts like fluorescence lines from Earth's atmosphere can result from improper background modeling. To check for these residuals, we re-extracted the spectrum from the brightest \nicer\ snapshot using the \texttt{nicerl3-spec} pipeline with the \texttt{bkgformat} option set to ``script'' instead of ``file'', allowing a proper background fit with the continuum. The residuals in the best-fit model were identical regardless of the method used, indicating that the observed spectral features are real and not due to the background. These residuals include two edges, at 0.32\,keV in \nicer\ data, and at 0.47\,keV in FXT data, likely due to calibration issues\footnote{See \url{https://heasarc.gsfc.nasa.gov/docs/nicer/analysis_threads/arf-rmf/}} and are not discussed. Instead, we detected intrinsic absorption edges from the source at energies of $\sim$0.63, $\sim$0.88 and $\sim1.0$\,keV, although they almost never appeared all together in a single spectrum. Two localised excesses at about 0.5--0.7\,keV and 0.9\,keV were visible in Epochs 1-14 and 1, 3-6, respectively.

We tested four models to describe the spectra: (a) a \texttt{BBodyrad} component plus a set of absorption edges and Gaussians; (b) a \texttt{BBodyrad} component plus a set of absorption edges and an \texttt{APEC} model, which describes the emission spectrum from a collisionally-ionized diffuse gas; (c) a NLTE stellar atmosphere model, obtained employing the grid 003 of the \texttt{TMAP}\footnote{\texttt{T\"ubingen} NLTE model-atmosphere package, \url{http://astro.uni-tuebingen.de/~rauch/TMAF/flux_HHeCNONeMgSiS_gen.html}} public repository \citep{Rauch2003} plus \texttt{APEC}; (d) a \texttt{SSS\_atm\_SMC} \citep{Suleimanov2024} \footnote{\texttt{SSS\_atm\_SMC} models the X-ray super-soft emission from a hot WD atmosphere with SMC abundances.} plus \texttt{APEC}. In all fits including \texttt{APEC}, the abundance parameter was unconstrained, so that we consistently kept it fixed to 0.1, i.e., the SMC abundances. Models (a) and (b) describe the spectra equally well. Acceptable fits are obtained with model (c) only for spectra corresponding to the decay phase of the outburst, i.e., from Epoch 11 on. At the peak of the outburst, the model falls short in describing the emission above 0.8 keV. The maximum temperature allowed by the \texttt{TMAP} model might be indeed too low to account for the relatively hard spectra in Epochs 1-10. For similar reasons, model (d) is also inadequate to describe our data, especially at the peak. In the following, we will mainly consider the results obtained with model (b), being more physically motivated than model (a) and the only model able to describe coherently our entire data set.

We consistently obtained $N_{\rm H}$ values within the range of (1--2)$\times$10$^{21}$\,cm$^{-2}$. This range is significantly larger than the expected foreground value for the SMC, which is only $\sim$6$\times$10$^{20}$\,cm$^{-2}$ \citep{Dickey1990}. The discrepancy between our obtained and the expected $N_{\rm H}$ values might indicate the presence of local absorption along the line of sight, as has been suggested for other SSS in the SMC \citep{coe20,kennea21}. From the blackbody normalization we estimated the radius of the blackbody emitting regions in each time-resolved spectrum, finding a range of about $3$-$6\times$10$^3$ km. The best-fit values obtained for both the continuum and the edges are listed in Tables \ref{tab:continuum} and \ref{tab:discrete}, respectively. The simultaneous evolution of the best-fit parameters can be observed in Figure\,\ref{fig:best-fit}. 

\subsection{Pre-outburst detections and post-outburst upper limits}
The upper limit on the net count rate measured in the last four stacked \swift/XRT observations and in the \xmm/EPIC-pn observation converts to unabsorbed bolometric fluxes of $<$6$\times$10$^{-12}$ and $<$7$\times$10$^{-13}$\,\flux, respectively, assuming the same parameters measured in the latest \nicer\ spectrum (see Table\,\ref{tab:continuum}). On the other hand, 
the count rate recorded in the 2002 \cxo\ observation, which has a higher signal-to-noise ratio compared to the 2006 observations, corresponds to an observed flux of $\approx3 \times 10^{-14}$\,\flux\ (0.3--8\,keV; computed using the \texttt{srcflux} tool, assuming an absorbed power-law model with an absorption column density of $2 \times 10^{21}$\,cm$^{-2}$ and a photon index in the range of 0.5 to 1.5). This results in a luminosity of $\approx1.5 \times 10^{34}$\,\lum\ (0.3-8\,keV). For the 2006 observations, assuming the same spectral shape as above and scaling based on the count rate, we estimate a luminosity that is about an order of magnitude lower, $\approx10^{33}$\,\lum.

\subsection{Searches for periodic signals}
\label{sec:periodicity}
To search for periodic signals, we used the data collected when the source was brightest and corrected the photon times of arrival to the barycenter of the Solar System. We considered data of the first two \nicer\ observations and the first EP/FXT observation, all restricted to the 0.3--2\,keV energy band to maximize the signal-to-noise ratio. \swift/UVOT data taken in event mode during the longest observation on June 17 were also searched. We then extracted power spectra from the unbinned event files from each dataset separately.
No statistically significant signals were detected in any of the data. The 3$\sigma$ upper limits on the pulsed fraction were calculated \citep{Israel1996}, resulting in: $<$4\% up to 2\,kHz in the \nicer\ data; between 10-16\% up to 250\,Hz in the EP/FXT data; between 7-10\% up to 50\,Hz in the \swift/UVOT data. For the \nicer\ data, which had the highest signal-to-noise ratio, we also employed Fourier-domain acceleration search techniques using \texttt{PRESTO}\footnote{\url{https://github.com/scottransom/presto}} \citep{Ransom2002} to search for signals up to 2\,kHz. This analysis was conducted on the entire dataset of the first two observations as well as on 300-s data chunks. No statistically significant signal was detected. These results are consistent with those reported by \cite{Jaisawal2024}. 

\section{Discussion and conclusions}
\label{sec:discussion}
In this paper, we have presented the EP/WXT discovery of an X-ray outburst from the BeWD binary \srcfirst\ \citep{antoniou2009} in the SMC and the follow-up X-ray observing campaign performed with EP/FXT, \nicer, \swift/XRT, and \xmm/EPIC. The source experienced a rapid increase in bolometric luminosity from $\sim$4$\times10^{33}$\,\lum\ before the onset of the outburst to $\sim$4$\times10^{38}$\,\lum\ at the outburst peak. A slower decay over one order of magnitude in luminosity was then observed in about six days, after which the system was no longer detectable by \nicer, likely due to being too faint, too soft, or both. An \xmm\,/EPIC-pn observation set a stringent upper limit of $<$5$\times$10$^{35}$\,\lum\ on the bolometric luminosity of the source 12 days after the last \nicer\ detection. This indicates a reduction in luminosity by at least three orders of magnitude during that period, marking the end of the outburst. We therefore conclude that the total duration of the outburst from \src\ ranged between 6 and 12 days, making it remarkably short. Throughout the entire outburst, the source consistently exhibited a soft spectral shape, with no emission detected beyond $\sim$1.5\,keV. However, the observed spectrum was not as soft as typical SSS, whose emission usually cuts off at $\sim$0.7 keV \citep[e.g.][]{Orio2022}, making this source somewhat peculiar. The spectra are well-modeled using a combination of a blackbody spectrum and an \texttt{apec} model. The source reached a blackbody temperature of $\sim$120\,eV in about one day since the EP/WXT detection and then exhibited a rapid decay to $\sim$50\,eV in less than a week. The blackbody radius showed an almost opposite trend, going from a minimum of $\sim$2700\,km coincident with the $kT$ peak to a maximum of $\sim$6700\,km after about 2 days and then plateauing for the remainder of the outburst. Similar trends have been reported in the SSS phase of several novae, analysed as well with blackbody models \citep[see, e.g.,][]{Page2015,Page2020}, although the peak temperatures were significantly lower than those seen in \src. A more remarkable similarity in both the peak temperatures and the spectral shape, i.e., not soft enough for a canonical SSS, can be spotted in the outbursting BeWD binary MAXI J0158-744 \citep{Li2012}. The initial temporary decrease in the temperature observed between epochs 1 and 3 could also be due to variability in the early SSS phase as reported by \cite{Osborne2010} or could be an artefact due to intercalibration issues between FXT, XRT and \nicer. We caution that blackbody models are only a crude approximation of the actual physics in SSS. Therefore, while the trends observed in the parameters are reliable, the specific values obtained for these parameters are less so. The attempt to model the thermal component with more physically reliable WD atmosphere models \citep{Rauch2003,Suleimanov2024} was instead not as successful. At the peak of the outburst, these models can not describe our data \citep[as in MAXI J0158-744,][]{Li2012}, possibly because of the upper limit on the WD temperature being too stringent and/or some of the model assumptions being not applicable to this source. Good fits are instead obtained in the decline phase of the outburst, a stage where possibly the emission of our source resembles more that of a hot compact white dwarf.

Almost all the spectra analysed in this paper showed a dense and intricate array of discrete spectral features in absorption and emission. The $\sim$0.63\,keV, $\sim$0.87\,keV and $\sim$0.98 keV edges can be associated\footnote{We mostly refer to \url{https://xdb.lbl.gov/Section1/Sec_1-8.html} for the identification of these lines.} with absorption edges from hydrogen-like species of N, O and Ne, respectively. Nitrogen, oxygen and neon features are often found in the X-ray spectra of novae and SSS \citep[see, e.g.][]{Ness2013,Drake2021,Bhargava2024} and they can be considered as a signature of a WD accretor. The additional emission lines are instead well-modelled by including in the fits a relatively cold, i.e., $kT_{\rm apec}\sim100-200$ eV, shocked gas emitting an \texttt{APEC} component. The presence of this component might suggest the presence of shocked ejecta around an emerging hot WD, as sometimes observed in novae with early turned-on SSS phases \citep[see, e.g.,][]{Page2020}. We conclude that the majority of the observed properties of the X-ray outburst of \src , with the exception of the rather hard spectral shape at the outburst peak and its incompatibility with WD atmosphere models, are in line with the SSS phase of a nova eruption. In this scenario, the observed X-ray outburst would be explained by thermonuclear burning on the surface of the WD, and not by the initial thermonuclear runaway phase of a small-amplitude nova, as previously suggested by \cite{Gaudin2024} based only on the first \swift\, observation of the source. However, the duration of the observed super-soft outburst, i.e., about a week, stands out with respect to most novae in the SSS phase, which typically lasts tens to hundreds of days (e.g. \citealt{Henze2010}, but see \citealt{Orio2023} for exceptions). The duration of an SSS phase depends primarily on the mass and chemical composition of the WD. WDs near the Chandrasekhar limit have shorter SSS phases since less accreted material is needed for a thermonuclear runaway. Conversely, higher hydrogen abundance in the accreted material slows the process \citep{Hachisu2006}. The combination of a $\sim$10 days duration, a peak effective temperature of $\sim$120 eV and a peak luminosity above $10^{38}$\,\lum, suggests the presence of a very massive object \citep{Wolf2013, Soraisam2016}. However, we caution that the existing literature may not be specifically tailored to the scenario of a WD accreting from a Be star. In addition, the presence of O and Ne features could be the signature of an Ne-O WD in the system, which is among the most massive WDs, typically exceeding 1.1 M$_\odot$ \citep{Taguchi2023}. Future high-resolution optical and/or X-ray spectra might shed light on the chemical composition of the WD and potentially confirm the accreting star in \src\ as a Ne-O WD. \\ 
It is noteworthy that in novae, the SSS phase typically begins after an optical-to-UV brightening phase caused by the expanding WD photosphere, which usually lasts for weeks \citep[e.g.][]{Darnley2016}.  In the case of \src, however, only a 0.5 magnitude increase in the optical band was observed six hours before the X-ray outburst \citep{Gaudin2024}, which would be unusual for a thermonuclear runaway-driven explosion. Similarly, a short-lived optical flare occurred just eleven hours before the X-ray outburst of the BeWD MAXI J0158-744 \citep{Li2012}. It is possible that the optical brightening in both that source and in \src\,  might have occurred but went undetected due to the dominant emission from the optically bright Be companion star \citep[as suggested by][]{Li2012}. Alternatively, an intrinsically short optical flare could be due the whole nova explosion being quick, possibly because of exceptionally low ejecta mass \citep{Morii2013}, as also suggested by \cite{Gaudin2024}.

\begin{figure*}
\centering
\includegraphics[scale=0.5]{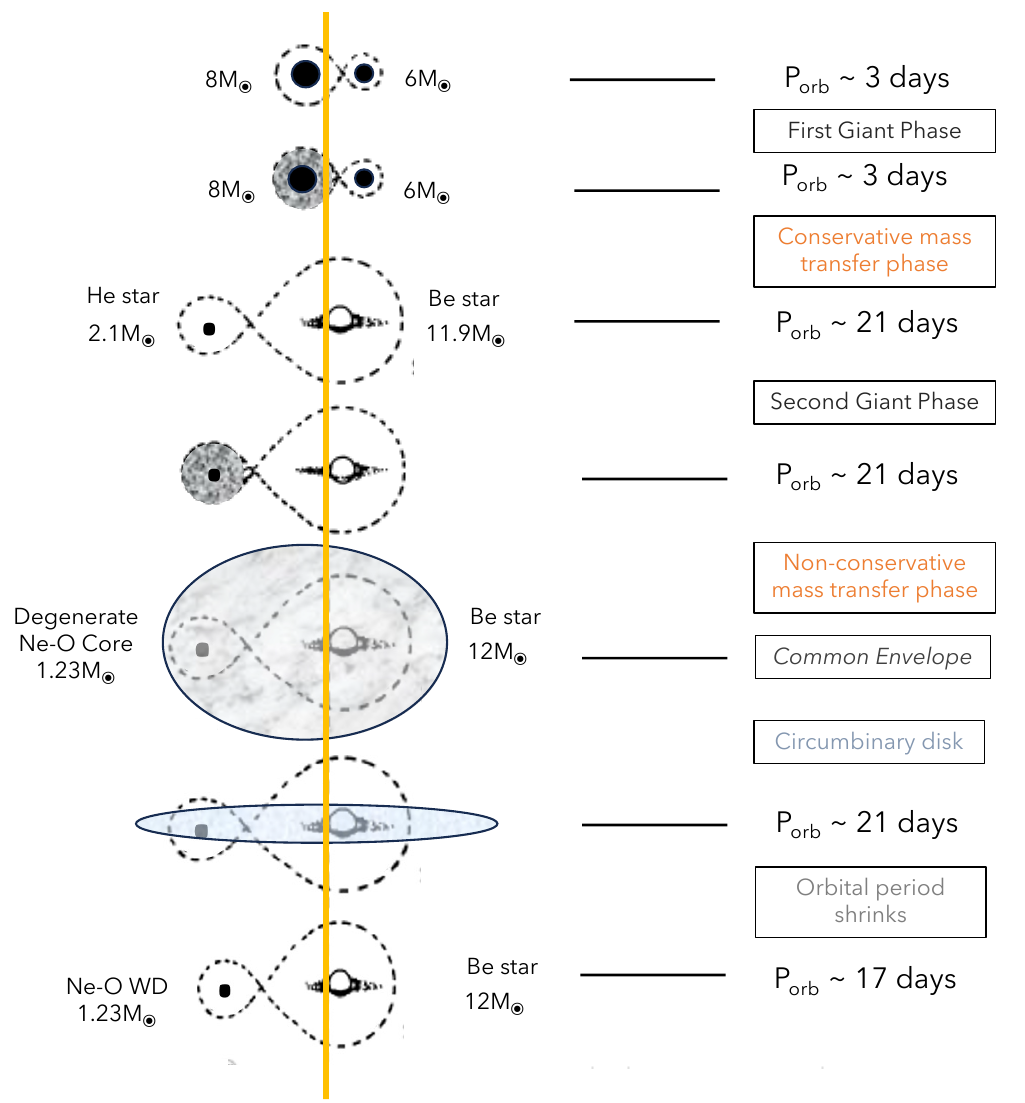}
\caption{Schematic evolutionary path of \src\, as described in the text. This is the lower-mass analogue of the evolutionary model by \citet{Habets1986a} for forming a Be- X-ray binary, here adapted for the case in which the accreting object is a massive Ne-O WD.} 
\label{Fig:model_evolution}
\end{figure*}

The ensemble of the X-ray spectral clues gathered in this work indicates that the compact object in \src\ is a massive, possibly Ne-O, WD. We therefore propose that \src\, is a BeWD binary. Such probably common but very elusive systems provide valuable insights into the evolution of binary stars, showcasing observable effects of the common envelope (CE) phase and mass transfer in massive star binaries. 
A binary system initially composed of two main-sequence stars of masses of about $8M_{\odot}$ and $6M_{\odot}$ with an orbital period of $\sim$3 days is a possible progenitor for \src. Following conservative mass transfer as computed by \citet{Gotberg2018}, the system evolves into a $2.1M_{\odot}$ He-star and a $11.9M_{\odot}$ Be main-sequence star with an orbital period of about 21 days. According to the computations by \citet{Woosley2019}, such a He-star terminates its evolution as a Ne-O WD with a mass of $1.23M_{\odot}$, confirming results by \citet{Habets1986b}. During the evolution of the He-star, mass transfer to the Be star resumes, but the mass transfer is non-conservative, due to the He-star’s convective envelope, which expands on a dynamical timescale, leading to the formation of a CE surrounding the Be star and the degenerate Ne-O core. When the envelope has expanded to the second Lagrangian point, it leads to the formation of a circumbinary disk with a mass of about $0.9M_{\odot}$. This causes the orbital period to decrease by a factor of about 0.8, resulting in a final orbital period of about 17 days (see section 4.3.8 of \citealt{Tauris2023} for orbital period effects of a circumbinary disk). The disk will finally be expelled from the system. We sketch the proposed evolutionary scenario for \src\, in Figure\,\ref{Fig:model_evolution}. It is noteworthy that, despite the above scenario being only one of the possible models for the secular evolution of the binary system, the expected age of the system fits very well with the \src\, age, i.e., of about 40 Myears \citep{Antoniou2019}. Indeed, the initial  8 M$_\odot$ star takes approximately $\sim$35 Myears before evolving into a system with a Be star of about 12-13 M$_{\odot}$, characterised by lifetimes of typically $\sim$20 Myears \citep{Schaller1992}. \\ 
If the nature of the source is confirmed, this work may represent one of the very few cases where a X-ray outburst in such systems has been thoroughly monitored for most of its duration, from rise to decay. The unprecedented monitoring capabilities of \ep\ in the soft X-ray range are particularly well-suited to discovering super-soft outbursts from these sources, offering the opportunity to address many unresolved questions about them. 

\begin{acknowledgments}
We thank the anonymous referee for their insightful and useful comments. This work is based on data obtained with \ep, a space mission supported by Strategic Priority Program on Space Science of Chinese Academy of Sciences, in collaboration with ESA, MPE and CNES (Grant No. XDA15310000), the Strategic Priority Research Program of the Chinese Academy of Sciences (Grant No. XDB0550200), and the National Key R\&D Program of China (2022YFF0711500). This work is also based on data obtained by: the Neil Gehrels \swift\ Observatory (a NASA/UK/ASI mission) supplied by the UK \swift\ Science Data Centre at the University of Leicester; \nicer, a 0.2--12\,keV X-ray telescope operating on the International Space Station, funded by NASA; \xmm, an ESA science mission with instruments and contributions directly funded by ESA Member States and NASA. This paper employs a list of Chandra datasets, obtained by the Chandra X-ray Observatory, contained in the Chandra Data Collection (CDC) with doi~\href{https://doi.org/10.25574/cdc.294}{doi:10.25574/cdc.294}. We thank N.~Schartel for approving Target of Opportunity observations with \xmm\ in the Director’s Discretionary Time and the \xmm\ Science Operation Center for scheduling and carrying out the observations. Similarly, we thank the \swift\ team for approving and performing our Target of Opportunity observations. 

We acknowledge the support by the National Natural Science Foundation of China (Grant Nos. 12321003, 12103065, 12203071, 12333004, 12373040, 12021003), the China Manned Space Project (Grant Nos. CMS-CSST-2021-A13, CMS-CSST-2021-B11), and the Youth Innovation Promotion Association of the Chinese Academy of Sciences. We acknowledge the data resources and technical support provided by the China National Astronomical Data Center, the Astronomical Science Data Center of the Chinese Academy of Sciences, and the Chinese Virtual Observatory. AM and NR are supported by the European Research Council (ERC) under the European Union’s Horizon 2020 research and innovation programme (ERC Consolidator Grant "MAGNESIA" No. 817661, PI: NR). FCZ is supported by a Ram\'on y Cajal fellowship (grant agreement RYC2021-030888-I). AM, FCZ and NR acknowledge support from grant SGR2021-01269 from the Catalan Government (PI: Graber/Rea). HY acknowledges support from China Scholarship Council (No.202310740002). NR acknowledges support from the ESA Science Faculty Visitor program to ESTEC (funding reference ESA-SCI-E-LE-054), where most of this work has been carried out. SG acknowledges the support of the CNES. We thank A. Zesas, D. De Martino, S. Scaringi, T. Maccarone and D. Buckley for fruitful discussion.
\end{acknowledgments}

\facilities{\emph{Einstein Probe}, \swift, \nicer, \xmm, \cxo}

\software{CIAO v4.16.0 \citep{fruscione2006}, fxtsoftware v1.05, HEASOFT v6.33.2 \citep{heasoft}, Matplotlib v3.9 \citep{hunter07}, NICERDAS v12, SAS v21.0.0	\citep{gabriel04}, XSPEC v12.14.0h \citep{arnaud96}}

\appendix
\restartappendixnumbering 
\renewcommand{\thetable}{A\arabic{table}}


\begin{table*}
    \centering
    \caption{Spectral analysis results using a black-body model for the continuum emission and \texttt{vapec}. The parameters that have been fixed are indicated in parenthesis. We consider MJD 60458.23 as reference time. Abundances in \texttt{vapec} fixed to 0.1.}
    \begin{tabular}{c c c c c c c c}
    \hline
    \hline
Time & nH & kT$_{bb}$ & R$_{bb}$ & kT$_{apec}$ & N$_{apec}$ & L$_{bol}$ & $\chi^2$ \T \B \\ 
  (d) & ($\times$10$^{22}$) & (eV) & ($\times$10$^3$ km) & (eV) & & ($\times$10$^{38}$ erg/s) & (d.o.f.) \T \B \\ 
  \hline
0.01 & 0.140$^{+0.007}_{-0.006}$ &100$^{+1}_{-2}$ &4.9$\pm$0.1 &240$\pm$4  & 0.28$\pm$0.04 &3.1$\pm$0.2 &411(287)\T \B \\  
0.43 & 0.180$^{+0.016}_{-0.017}$ &110$^{+4}_{-3}$ &4.0$^{+0.7}_{-0.5}$ &230$^{+16}_{-18}$  & (0.30) &3.90$\pm$0.07 &87(73)\T \B \\   
0.55 & 0.140$\pm$0.003 &120$\pm$2 &2.8$^{+0.2}_{-0.1}$ &210$\pm$6  & 0.29$\pm$3 &2.80$\pm$0.02 &98(90)\T \B \\  
0.61 & 0.140$^{+0.004}_{-0.003}$ &120$\pm$2 &2.9$^{+0.2}_{-0.1}$ &210$\pm$6 & 0.27$^{+0.03}_{-0.02}$ &2.70$\pm$0.02 &103(90)\T \B \\  
0.67 & 0.140$\pm$0.003 &120$\pm$2 &2.9$^{+0.2}_{-0.1}$ &210$^{+6}_{-7}$  & 0.25$\pm$0.03 &2.60$\pm$0.02 &107(90)\T \B \\  
0.74 & 0.140$\pm$0.004 &110$\pm$2 &3.2$\pm$0.2 &190$\pm$8  & 0.25$\pm$0.03 &2.90$\pm$0.02 &124(90)\T \B \\  
1.25 & 0.170$\pm$0.006 &90$\pm$2 &5.5$^{+0.5}_{-0.4}$ &170$\pm$5  & 0.30$\pm0.02$ &3.80$\pm$0.03 &100(79)\T \B \\  
1.45 & 0.150$^{+0.006}_{-0.005}$ &94$\pm$2 &4.5$^{+0.5}_{-0.4}$ &170$\pm$8  & 0.25$\pm$0.03 &2.90$\pm$0.02 &87(69)\T \B \\  
1.51 & 0.150$\pm$0.006 &92$\pm$2 &4.7$^{+0.5}_{-0.4}$ &150$\pm$5  & 0.30$^{+0.06}_{-0.05}$ &3.30$\pm$0.03 &84(65)\T \B \\  
1.58 & 0.160$^{+0.008}_{-0.007}$ &89$\pm$2 &5.3$^{+0.7}_{-0.5}$ &160$^{+9}_{-10}$  & 0.22$\pm$0.04 &3.00$\pm$0.02 &85(64)\T \B \\  
1.64 & 0.160$\pm$0.008 &87$^{+3}_{-2}$ &5.5$^{+0.7}_{-0.6}$ &160$^{+11}_{-10}$  & 0.18$^{+0.04}_{-0.03}$ &2.80$\pm$0.02 &79(63)\T \B \\ 
1.71 & 0.160$^{+0.008}_{-0.007}$ &86$\pm$2 &5.4$^{+0.7}_{-0.5}$ &140$^{+10}_{-13}$  & 0.24$^{+0.10}_{-0.06}$ &3.20$\pm$0.03 &71(63)\T \B \\  
2.03 & 0.140$^{+0.007}_{-0.006}$ &80$\pm$3 & 5.0$^{+0.9}_{-0.6}$ &(140) & 0.15$\pm$0.03 &2.00$\pm$0.02 &58(37)\T \B \\  
2.16 & 0.140$\pm$0.008 &81$\pm$3 &4.8$^{+0.8}_{-0.6}$ &(140) & 0.06$\pm$0.02 &1.50$\pm$0.01 &39(42)\T \B \\  
2.29 & 0.140$^{+0.011}_{-0.010}$ &78$\pm$4 &4.9$^{+1.3}_{-0.8}$ & (140) & 0.06$^{+0.03}_{-0.02}$ &1.34$\pm$0.02 &40(41)\T \B \\  
2.42 & 0.140$^{+0.011}_{-0.010}$ &77$\pm$4 &4.8$^{+1.2}_{-0.7}$ &(140) & 0.04$\pm$0.02 &1.20$\pm$0.01 &30(42)\T \B \\  
3.06 & 0.130$\pm$0.010 &74$^{+3}_{-2}$ &3.7$^{+0.6}_{-0.5}$ &- & - & 0.46$\pm$0.01 &32(40)\T \B \\  
3.13 & 0.110$\pm$0.010 &76$\pm$2 &3.1$\pm$0.3 &- & - & 0.35$\pm$0.01 &35(42)\T \B \\  
4.29 & (0.130) &56$\pm$2 &5.6$^{+0.6}_{-0.5}$ &- & - & 0.38$\pm$0.02 &24(25)\T \B \\  
5.06 & (0.130) &53$\pm$1 &5.2$^{+0.5}_{-0.4}$ &- & - & 0.25$\pm$0.01 &35(26)\T \B \\  
5.96 & (0.130) &50$\pm$2 &5.1$^{+1.0}_{-0.7}$ &- & - & 0.19$\pm$0.02 &16(21)\T \B \\ 

\hline
    \end{tabular}
    \label{tab:continuum}
\end{table*}



\begin{table*}
    \centering
    \footnotesize
    \caption{Spectral analysis of the absorption edges included in the fits. The parameters that have been fixed are indicated in parenthesis. We consider MJD 60458.23 as reference time.}
    \begin{tabular}{c c c c c c c c c }
    \hline
    \hline
    \multicolumn{9}{c}{Absorption edges} \\
    \hline
Time & E$_{e,1}$ & $\tau_{e,1}$ & E$_{e,2}$ & $\tau_{e,2}$ & E$_{e,3}$ & $	\tau_{e,3}$ & E$_{e,4}$ & $\tau_{e,4}$ \T \B \\  
(d) & (keV) & - & (keV) & - & (keV) & - & (keV) & - \T \B \\ 
\hline
0.01 & 0.370$\pm$0.006& (1.0)  & - & - & 0.910$\pm$0.004 & 1.0$\pm$0.6& - & - \T \B \\  
0.43 & - & - & - & - & 0.870$\pm$0.001& 3.0$^{+2.0}_{-0.8}$& - & - \T \B \\   
0.55 & 0.330$\pm$0.005& (1.0)  & - & - & 0.870$\pm$0.003& 1.0$\pm$0.7& 0.990$^{+0.006}_{-0.005}$& 1.20$\pm$0.08  \T \B\\  
0.61 & 0.330$\pm$0.005& (1.0)  & - & - & 0.870$\pm$0.003& 1.2$\pm$0.1& 0.990$\pm$0.006& 1.20$\pm$0.08  \T \B\\  
0.67 & 0.330$\pm$0.005& (1.0)  & - & - & 0.870$\pm$0.004& 1.2$\pm$0.1& 0.980$\pm$0.007& 1.20$\pm$0.10  \T \B\\  
0.74 & 0.340$^{+0.005}_{-0.006}$& (1.0)  & - & - & 0.870$\pm$0.005& 1.1$\pm$0.1& 0.980$\pm$0.009& 1.20$\pm$0.12  \T \B\\  
1.25 & 0.340$^{+0.005}_{-0.006}$& (1.0)  & - & - & 0.870$\pm$0.002& 2.8$^{+0.3}_{-0.2}$& - & - \T \B \\  
1.45 & 0.330$\pm$0.007& (1.0)  & 0.670$\pm$0.008& 0.39$^{+0.08}_{-0.07}$& 0.850$\pm$0.003& 3.1$^{+0.4}_{-0.3}$& - & - \T \B \\  
1.51 & 0.340$\pm$0.005& (1.0)  & 0.680$\pm$0.008& 0.45$\pm$0.01& 0.850$\pm$0.003& 2.9$\pm$0.3& - & - \T \B \\  
1.58 & 0.340$\pm$0.006& (1.0)  & 0.680$\pm$0.006& 0.62$\pm$0.08& 0.850$^{+0.003}_{-0.004}$& 3.3$^{+0.5}_{-0.4}$& - & - \T \B \\  
1.64 & 0.340$\pm$0.006& (1.0)  & 0.670$\pm$0.005& 0.84$^{+0.10}_{-0.09}$& 0.840$^{+0.004}_{-0.005}$  & 3.4$^{+0.8}_{-0.5}$& - & - \T \B \\  
1.71 & 0.340$\pm$0.006& (1.0)  & 0.670$\pm$0.004& 1.10$^{+0.08}_{-0.07}$& 0.840$\pm$0.003& 3.4$^{+0.6}_{-0.4}$& - & - \T \B \\  
2.03 & 0.300$^{+0.013}_{-0.030}$& (1.0)  & 0.670$\pm$0.003& 1.80$^{+0.12}_{-0.12}$& - & - & - & - \T \B \\  
2.16 & 0.320$^{+0.007}_{-0.009}$& (1.0)  & 0.660$\pm$0.002& 2.50$\pm$0.11& - & - & - & - \T \B \\  
2.29 & 0.310$^{+0.009}_{-0.012}$& (1.0)  & 0.660$\pm$0.003& 2.60$^{+0.16}_{-0.15}$& - & - & - & - \T \B \\  
2.42 & 0.320$^{+0.008}_{-0.009}$& (1.0)  & 0.660$\pm$0.002& 2.90$^{+0.18}_{-0.16}$& - & - & - & - \T \B \\  
3.06 & (0.300) & (1.0)  & 0.650$\pm$0.004& 2.60$\pm$0.20& - & - & - & - \T \B \\  
3.13 & (0.300) & (1.0)  & 0.660$\pm$0.003& 2.90$\pm$0.20& - & - & - & - \T \B \\  
4.29 & (0.300)& (1.0)  & - & - & - & - & - & - \T \B \\   
\hline
    \end{tabular}
    \label{tab:discrete}
\end{table*}


\bibliography{refs}{}
\bibliographystyle{aasjournal}

\end{document}